\documentclass[epj]{webofc}
\usepackage[varg]{txfonts}   % Web of Conferences font
\usepackage{graphicx,color} % for figures
\usepackage{bm}       % bold math 
\usepackage{amsmath}  
\usepackage{amssymb}
\usepackage{calc}
\usepackage{textcomp}
\usepackage{graphicx}
\usepackage{enumerate}
\woctitle{21st International Conference on Few-Body Problems in Physics}
\begin{document}
\title{Electric dipole polarizability: from few- to many-body systems}
\author{Mirko Miorelli\inst{1,2}\fnsep\thanks{\email{mmiorelli@triumf.ca}} \and
        Sonia Bacca\inst{1,3}\fnsep\thanks{\email{bacca@triumf.ca}} \and
        Nir Barnea\inst{4}\and
	 Gaute Hagen\inst{5,6}\and
	 Giuseppina Orlandini\inst{7,8}\and
	 Thomas Papenbrock\inst{5,6}
        % etc.
}

\institute{TRIUMF, 4004 Wesbrook Mall,
Vancouver, British Columbia, Canada V6T 2A3
\and
           Department of Physics and Astronomy, University of British Columbia,
Vancouver, British Columbia, Canada V6T 1Z4
\and
           Department of Physics and Astronomy, University of Manitoba,
Winnipeg, Manitoba, Canada R3T 2N2
\and
	Racah Institute of Physics, Hebrew University,
91904 Jerusalem, Israel
\and
	Physics Division, Oak Ridge National Laboratory,
Oak Ridge, Tennessee 37831, USA
\and
	Department of Physics and Astronomy, University of Tennessee,
Knoxville, Tennessee 37996, USA
\and
	Dipartimento di Fisica, Universit\'a di Trento, Via Sommarive 14, I-38123 Trento, Italy
\and
	Istituto Nazionale di Fisica Nucleare, TIFPA,
Via Sommarive 14, I-38123 Trento, Italy
          }

\abstract{
We review the Lorentz integral transform coupled-cluster method for the calculation of the electric dipole polarizability. We benchmark our results with exact hyperspherical harmonics calculations for \textsuperscript{4}He and then we move to a heavier nucleus studying \textsuperscript{16}O. We observe that the implemented chiral nucleon-nucleon interaction at next-to-next-to-next-to-leading order underestimates the electric dipole polarizability.}
\maketitle
\section{Introduction}
\label{intro}
The electric dipole polarizability is a fundamental quantity to understand nuclear dynamics and is related to the response of the nucleus to an external electric field. Typically, it is measured via photo-absorption reactions \cite{Ahrens75},  Compton scattering~\cite{compton} or hadronic processes, such as $(p,p')$ reactions \cite{Tamii14}. Ab-initio calculations of this observable were traditionally available only for very light nuclei. It is our goal to extend such studies to the medium mass regime using coupled-cluster theory.

\section{Review of the method}
\label{sec-method}

The electric dipole polarizability can be calculated from the dipole response function $R(\omega)$ as an inverse energy  weighted sum rule 
\begin{equation}\label{pol_sumrule}
\alpha_D = 2\alpha\int_{\omega_{th}}^\infty \frac{R(\omega)}{\omega}d\omega.
\end{equation}
In the coupled-cluster formalism \cite{LITCC} the dipole response function is
\begin{equation}\begin{split}\label{respfunc}
R(\omega) = \sum_n &\langle 0_L|e^{-\hat{T}}{\Theta}^\dag e^{\hat{T }}|n_R\rangle\langle n_L|e^{-\hat{T }}{\Theta}e^{\hat{T }}|0_R\rangle\delta(E_n - E_0 - \omega),
\end{split}\end{equation}
where $\hat{\Theta}$ is the dipole excitation operator,
 $\hat{T}$ is the cluster operator defined in coupled-cluster theory \cite{shavittbartlett2009}, and $\langle 0_L|$, $|0_R\rangle$ ($\langle n_L|$, $|n_R\rangle$) are the left and right reference ground states (excited states), respectively. 
Using Eq.~(\ref{pol_sumrule}) we can then rewrite the electric dipole polarizability as 
\begin{equation}\label{pol_lit2}
\alpha_D= 2\alpha\langle 0_L|\overline{\Theta}^\dag|\tilde{\Psi}_R\rangle,
\end{equation}
where $\overline{\Theta} = e^{-\hat{T }}{\Theta}e^{\hat{T }}$ is the similarity transformed operator and  $\tilde{\Psi}_R$ is the solution of the Schr\"odinger-like equation
\begin{equation}\label{LITCC_eq}
(\overline{H} - \Delta E_0)|\tilde{\Psi}_R\rangle = \overline{\Theta}|0_R\rangle,
\end{equation}
where $\Delta E_0$ is the correlation energy of the ground state. This equation is obtained using the Lorentz integral transform method~\cite{LIT} in the coupled-cluster formulation~\cite{Bacca13,LITCC}.
Similarly to what was done in \cite{Nevo14, Goerke12}, the polarizability  can be expressed as a continued fraction of the Lanczos coefficients~\cite{Miorelli15}
\begin{equation}\label{pol_lit4}
\alpha_D = 2\alpha \langle 0_L|\overline{\Theta}^\dag\overline{\Theta}|0_R\rangle\left\{\frac{1}{a_0 - \frac{b_0^2}{a_1 - \frac{b_1^2}{a_2 - \cdots}}}\right\}.
\end{equation}
If one is able to calculate the response function of Eq.~(\ref{respfunc}), then the polarizability could be obtained using the definition in Eq.~(\ref{pol_sumrule}). However, using Eq.~(\ref{pol_lit4}) allows us to calculate the polarizability directly from the Lanczos coefficients, thus avoiding further uncertainties coming from the inversion of the transform. Furthermore, due to the nature of the continued fraction, convergence is reached with less than 100 Lanczos steps.
Eq.~(\ref{pol_lit4}) is an exact result, free of any approximation, if the cluster expansion is performed up to $A$-$particle$-$A$-$hole$ excitations, $A$ being the total number of nucleons. However, in practical calculations the cluster expansion is truncated since a full expansion is not feasible due to the very demanding computational cost. In the results we present here we truncated the cluster operators at the singles-and-doubles excitations level and dub this approximation scheme LIT-CCSD.

\section{Results and Conclusions}
\label{sec:results}
\subsection{Benchmark in \textsuperscript{4}He}
\label{subsec: 4He}
While the LIT-CCSD method has been already benchmarked with the effective interaction hyperspherical harmonic (EIHH) method \cite{Barnea00} for the dipole response function \cite{LITCC}, here we show a benchmark for the electric dipole polarizability calculated using Eq.~(\ref{pol_lit4}) with  a chiral nucleon-nucleon force derived at next-to-next-to-next-to-next-to-leading order~\cite{Entem03}.\\
\begin{figure}
\centering
\begin{minipage}{.45\linewidth}
  \includegraphics[width=\linewidth]{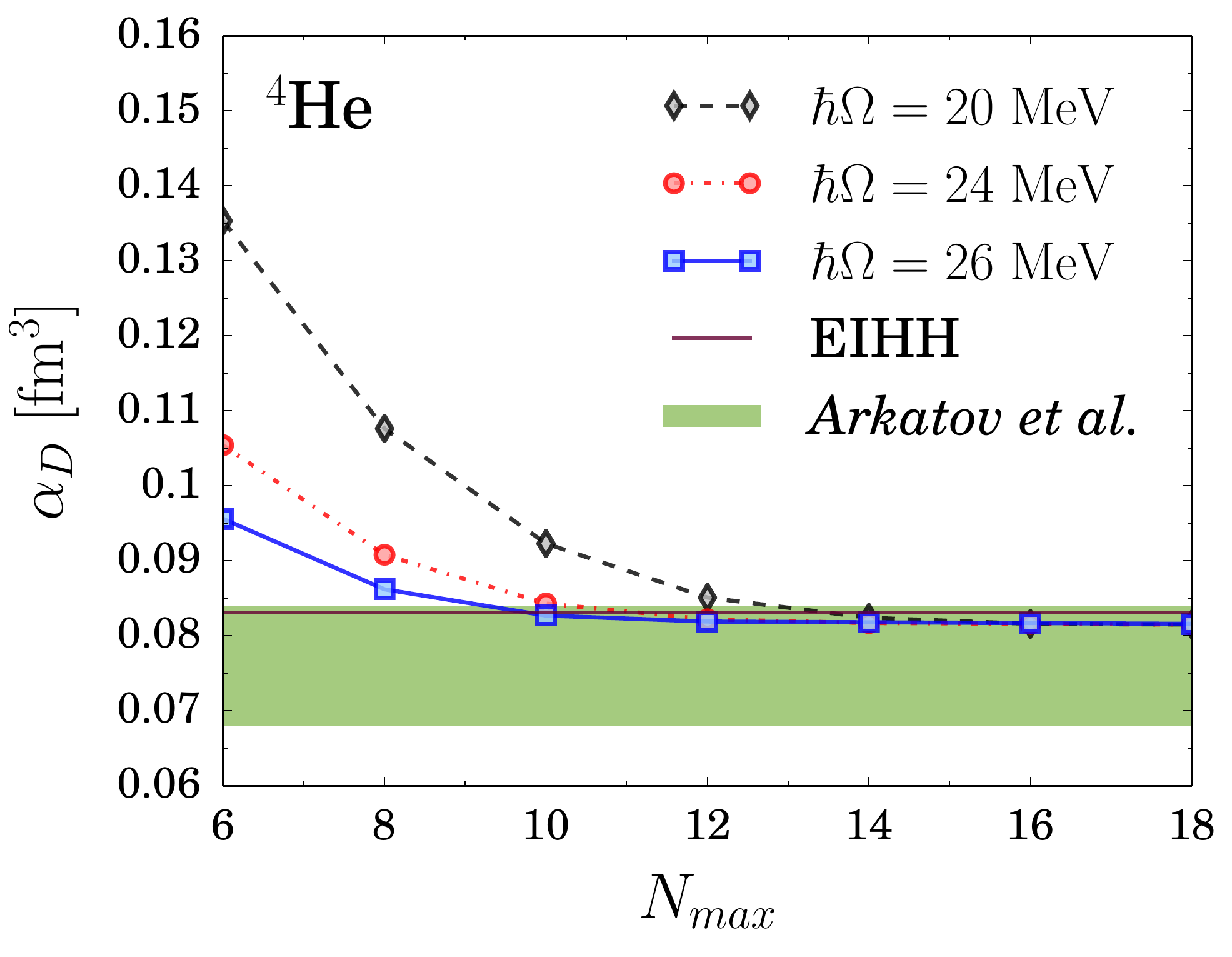}
  \caption{The electric dipole polarizability in $^{4}$He as a function of the model space size $N_{max}$ for different values of the harmonic oscillator frequency $\hbar\Omega$ (black, red, blue). The converged value at $N_{max}=18$ is compared with the result obtained from the EIHH method (brown) and the experimental data from \cite{Arkatov74, Arkatov80, Friar77, Pachucki07} (green band).}
  \label{fig:fig_pol_conv_4He}
\end{minipage}
\hspace{.05\linewidth}
\begin{minipage}{.45\linewidth}
  \includegraphics[width=\linewidth]{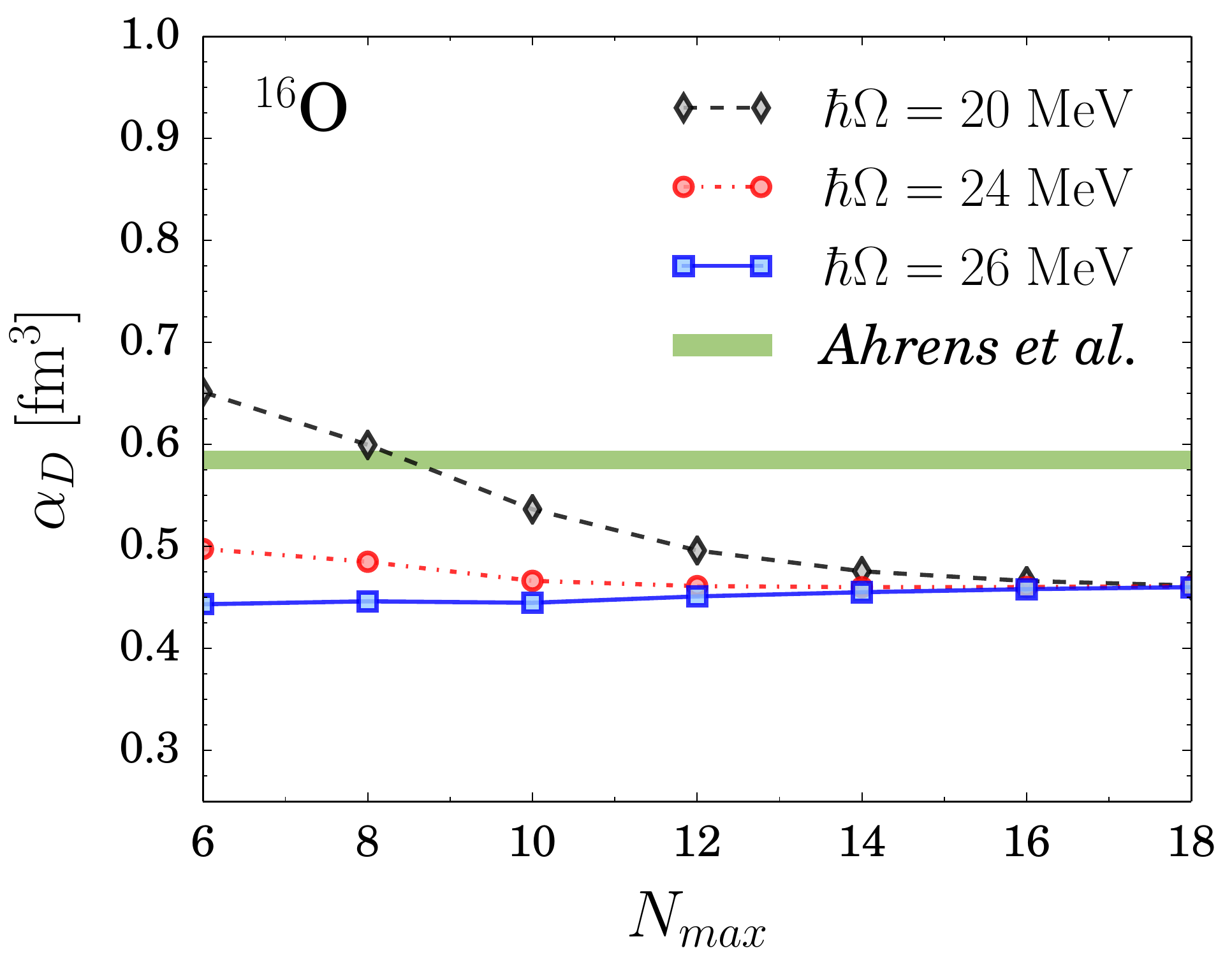}
  \caption{Comparison of the electric dipole polarizability in $^{16}$O as a function of the model space size $N_{max}$ for three different values of $\hbar\Omega=20,24$ and $26\ \rm{MeV}$ (black, red, blue) with the experimental data from Ahrens \textit{et al.} \cite{Ahrens75} (green).}
  \label{fig:fig_pol_conv_16O}
\end{minipage}
\end{figure}
In Figure \ref{fig:fig_pol_conv_4He}, we show the electric dipole polarizability of \textsuperscript{4}He as a function of the model space size $N_{max}$. The green band represents the combined experimental values from \cite{Friar77,Pachucki07} which have been extracted by fitting to experimental photo-absorption data \cite{Arkatov74, Arkatov80}. The continuous line (brown) is the polarizability obtained with the EIHH method. The lines with scatter-symbols are the results from the LIT-CCSD method obtained for different values of the frequency $\hbar\Omega$ of the harmonic oscillator basis used to expand the potential.
As expected, the polarizability is independent of the oscillator frequency as one increases $N_{max}$, and converges rapidly to a value $\alpha_D = 0.0815\ \rm{fm^3}$, which compares fairly well with the EIHH \cite{Goerke12} result of $\alpha_D = 0.0831\ \rm{fm^3}$ and $\alpha_D = 0.0822(5)\ \rm{fm^3}$ obtained using the no-core-shell-model with the same interaction \cite{Stetcu09}.
Moreover, the convergence with respect to the number of Lanczos coefficients used in Eq.~(\ref{pol_lit4}) is much faster than the one observed for the response function in \cite{LITCC}. Already with 15 Lanczos coefficients the error with respect to the converged value is $0.3\%$, which shows the advantage of using directly Eq.~(\ref{pol_lit4}). The small difference between the LIT-CCSD and EIHH results can thus be attributed to the truncation to singles-and-doubles excitations only.

\subsection{Application to \textsuperscript{16}O}
\label{subsec: 16O}
The very nice agreement in the benchmark for the $^4$He polarizability suggests that we can  extend the calculation of $\alpha_D$ via this method to heavier nuclei. We then consider \textsuperscript{16}O, whose response function has been already studied previously \cite{LITCC, Bacca13}. In Figure~\ref{fig:fig_pol_conv_16O}, we study the convergence of the polarizability as a function of the model space size. Again, the convergence is very fast and the final value $\alpha_D = 0.461\ \rm{fm^3}$ is independent of the underlying harmonic oscillator frequency. However, the converged  value  is rather low if compared to the experimental one of $\alpha_D^{exp} = 0.585(9)$ \cite{Ahrens75}. This trend is observed also in heavier nuclei such as \textsuperscript{40}Ca and \textsuperscript{48}Ca \cite{Miorelli15, LITCC}, and the discrepancy is even larger when the mass number is increased. As shown in Figure~\ref{fig:fig_pol_conv_4He}, the effect of the cluster expansion truncation is tiny and, because of the size extensive nature of coupled-cluster theory, we expect the effect to be very small in heavier systems as well. Discrepancies between the calculated values and the experimental data are mainly attributed to deficiencies of the employed Hamiltonian. 

\subsection{Conclusions}
\label{subsec:conclusions}
We used the LIT-CCSD method to extend the calculation of the electric dipole polarizability from few- to many-body nuclei using nucleon-nucleon chiral interactions. We have shown the reliability of the method with benchmarks in \textsuperscript{4}He where both EIHH and LIT-CCSD calculations and the experimental data are in agreement. Moving to heavier systems, \textit{e.g.}, \textsuperscript{16}O, we observe  disagreement between calculated and measured values. This points to the inadequacy of chiral interactions at the two-body level in accounting for the dipole polarizability in heavier nuclei, which we also observed for the radii~\cite{LITCC, Miorelli15}. The use of newly developed Hamiltonians with three-nucleon forces~\cite{Ekstroem15} calibrated on radii of finite nuclei can help solving this problem. Work in this direction is underway.

\begin{acknowledgement}
This work was supported in parts by the Natural Sciences and Engineering Research Council (NSERC), the National Research Council of Canada, the US-Israel Binational Science Foundation (Grant No.~2012212), the Pazy Foundation, the U.S. Department of Energy (Oak Ridge National Laboratory), under Grant Nos. DEFG02-96ER40963 (University of Tennessee), de-sc0008499 (NUCLEI Sci-DAC collaboration), and the Field Work Proposal ERKBP57.
\end{acknowledgement}

%
% BibTeX or Biber users please use (the style is already called in the class, ensure that the "woc.bst" style is in your local directory)
% \bibliography{name or your bibliography database}

\begin{thebibliography}{}
\bibitem{Ahrens75}
J. Ahrens, H. Borchert, K.H. Czock, H.B. Eppler, H. Gimm, H. Gundrum, M. Kröning, P. Riehn, G. Sita Ram, A. Zieger, and B. Ziegler, Nuclear Physics A \textbf{251} 479-492 (1975)
\bibitem{compton}
 H.W. Griesshammer, J.A.~McGovern, D.R.~Phillips, G.~Feldman, Prog.~Part.~Nucl.~Phys., {\bf 67}, 841-897 (2012)
\bibitem{Tamii14}
 A. Tamii, P. von Neumann-Cosel, and I. Poltoratska, Eur. Phys. J. A \textbf{50} 28 (2014) 
\bibitem{LITCC}
S. Bacca, N. Barnea, G. Hagen, M. Miorelli, G. Orlandini and T. Papenbrock, Phys. Rev. C \textbf{90}, 064619 (2014)
\bibitem{shavittbartlett2009}
I. Shavitt, and R. J. Bartlett, \textit{Many-body Methods in Chemistry and Physics} (Cambridge University Press, Cambridge UK, 2009) 255-256
\bibitem{LIT}
Efros, V.D.,  Leidemann, W.,  Orlandini, G.:
%Response functions from integral transforms with a Lorentz kernel. 
Phys. Lett. B  338, 130 (1994)
%\bibitem{Stanton93}
%J. F. Stanton, and R. J. Bartlett, The Journal of Chemical Physics \textbf{98}, 7029-7039 (1993)
\bibitem{Bacca13}
S. Bacca, N. Barnea, G. Hagen, G. Orlandini and T. Papenbrock, Phys. Rev. Lett. \textbf{111}, 122502 (2013)
\bibitem{Nevo14}
N. Nevo Dinur, J. Chen, S. Bacca and N. Barnea, Phys. Rev. C \textbf{89}, 064317 (2014)
\bibitem{Goerke12}
R. Goerke, S. Bacca, and N. Barnea, Phys. Rev. C \textbf{86} 064316 (2012)
\bibitem{Miorelli15}
M. Miorelli, S. Bacca, N. Barnea, G. Hagen, G. Orlandini and T. Papenbrock, in preparation.
\bibitem{Barnea00}
N. Barnea, W. Leidemann, and G. Orlandini, Phys. Rev. C \textbf{61}, 054001 (2000)
\bibitem{Entem03}
D. R. Entem, and R. Machleidt, Phys. Rev. C \textbf{68} 041001 (2003)
\bibitem{Friar77}
J. L. Friar, Phys. Rev. C \textbf{16} 1540-1548 (1977)
\bibitem{Pachucki07}
K. Pachucki, and A. M. Moro, Phys. Rev. A \textbf{75} 032521 (2007)
\bibitem{Arkatov74}
Y. M. Arkatov, P. I. Vatset, V. I. Voloshuchuk, V. A. Zolenko, I. M. Prokhorets, and V. I. Chimil’, Sov. J. Nucl. Phys. \textbf{19} 598 (1980)
\bibitem{Arkatov80}
Y. M. Arkatov, P. I. Vatset, V. I. Voloshuchuk, V. A. Zolenko, and I. M. Prokhorets, Sov. J. Nucl. Phys. \textbf{31} 726 (1980)
%\bibitem{Navratil00}
%P. Navr\'atil,  G. P. Kamuntavi\v{c}ius, and B. R. Barrett, Phys. Rev. C \textbf{61} 044001 (2000)
\bibitem{Stetcu09}
I. Stetcu, S. Quaglioni, J. L.  Friar, A. C. Hayes,  and P. Navr\'atil Phys. Rev. C \textbf{79} 064001 (2009)
\bibitem{Ekstroem15}
A. Ekstr\"om, G. R. Jansen, K. A. Wendt, G. Hagen, T. Papenbrock, B. D. Carlsson, C. Forssén, M. Hjorth-Jensen, P. Navrátil, and W. Nazarewic, Phys. Rev. C \textbf{91} 051301(R) (2015)
\end{thebibliography}

\end{document}